\documentclass[prl,aps,twocolumn,nofootinbib]{revtex4}

\usepackage{color}
\def\gr{$\gamma$-ray}

\usepackage[utf8]{inputenc}
\usepackage{graphicx}

\begin{document}
\title{Neutrino signal from Cygnus region of the Milky Way}
\author{Andrii Neronov$^{1,2}$, Dmitri Semikoz$^1$, Denys Savchenko$^{1,3,4}$}

\affiliation{$^{1}$Universit\'e Paris Cit\'e, CNRS, Astroparticule et Cosmologie, 
F-75013 Paris, France}

\affiliation{$^{2}$Laboratory of Astrophysics, \'Ecole Polytechnique F\'ed\'erale de Lausanne, CH-1015 Lausanne, Switzerland}

\affiliation{$^{3}$Bogolyubov Institute for Theoretical Physics of the NAS of Ukraine, 03143 Kyiv, Ukraine}

\affiliation{$^{4}$Kyiv Academic University, 03142 Kyiv, Ukraine}

\begin{abstract}
Interactions of cosmic ray protons and nuclei in their sources and in the interstellar medium produce ``hadronic"  \gr\ emission. Gamma-rays can also be of ``leptonic" origin, i.e. originating from high-energy electrons accelerated together with protons. It is difficult to distinguish between hadronic and leptonic emission mechanisms based on \gr\ data alone. This can be done via detection of neutrinos, because only hadronic processes lead to neutrino production. We use publicly available ten-year IceCube neutrino telescope dataset    to demonstrate the hadronic nature of high-energy emission from the direction of Cygnus region of the Milky Way. We find a $3\sigma$ excess of neutrino events from an extended Cygnus Cocoon, with the flux  comparable to the flux of \gr s in the multi-TeV energy range seen by HAWC and LHAASO telescopes. 
\end{abstract}

\maketitle

\textbf{Introduction.} The bulk of cosmic rays penetrating the Earth atmosphere is composed of high-energy protons and atomic nuclei coming from yet uncertain astronomical sources in the Milky Way galaxy. Astronomical objects produced in supernova explosions and superbubbles of star formation are considered as candidate cosmic ray source classes  \cite{cr_models}. Firm identification of the cosmic ray sources can be done via observations of multi-messenger neutrino and \gr s produced in cosmic ray interactions. 

IceCube telescope has found an evidence for  the neutrino signal from the Milky Way \cite{IceCube:2023ame,neronov14,neronov_diffuse,Neronov:2023dgm}, but has not yet located individual cosmic ray sources in the Galaxy. Contrary to the neutrino signal, high-energy \gr\ flux from cosmic ray proton and nuclei interactions has perhaps been already detected from multiple known Galactic sources. However, \gr s  can be produced not only in interactions of the cosmic ray protons and nuclei, but also by high-energy electrons. The only way to unambiguously distinguish ``hadronic" (proton or nuclei-powered) and ``leptonic" (electron-powered) emission is to detect the neutrino flux accompanying the \gr\ flux from proton and nuclei interactions.  

Previous $\gamma$-ray based estimate of neutrino flux from the Galactic sources  provides a prediction that the most promising candidate neutrino source in the Northern hemisphere is Cygnus region of the Milky Way disk, detectable at more than $3\sigma$ significance level in decade-scale exposure  \cite{tchernin13}.  This active star formation region \cite{cygnus_x} hosts multiple massive star associations, supernova remnants, pulsar wind nebulae \cite{cygnus_projection} and a hard \gr\ spectrum diffuse emission region Cygnus Cocoon \cite{cocoon_fermi,hegra,milagro,argo,hawc}. 

\begin{figure}
    \includegraphics[width=0.9\linewidth]{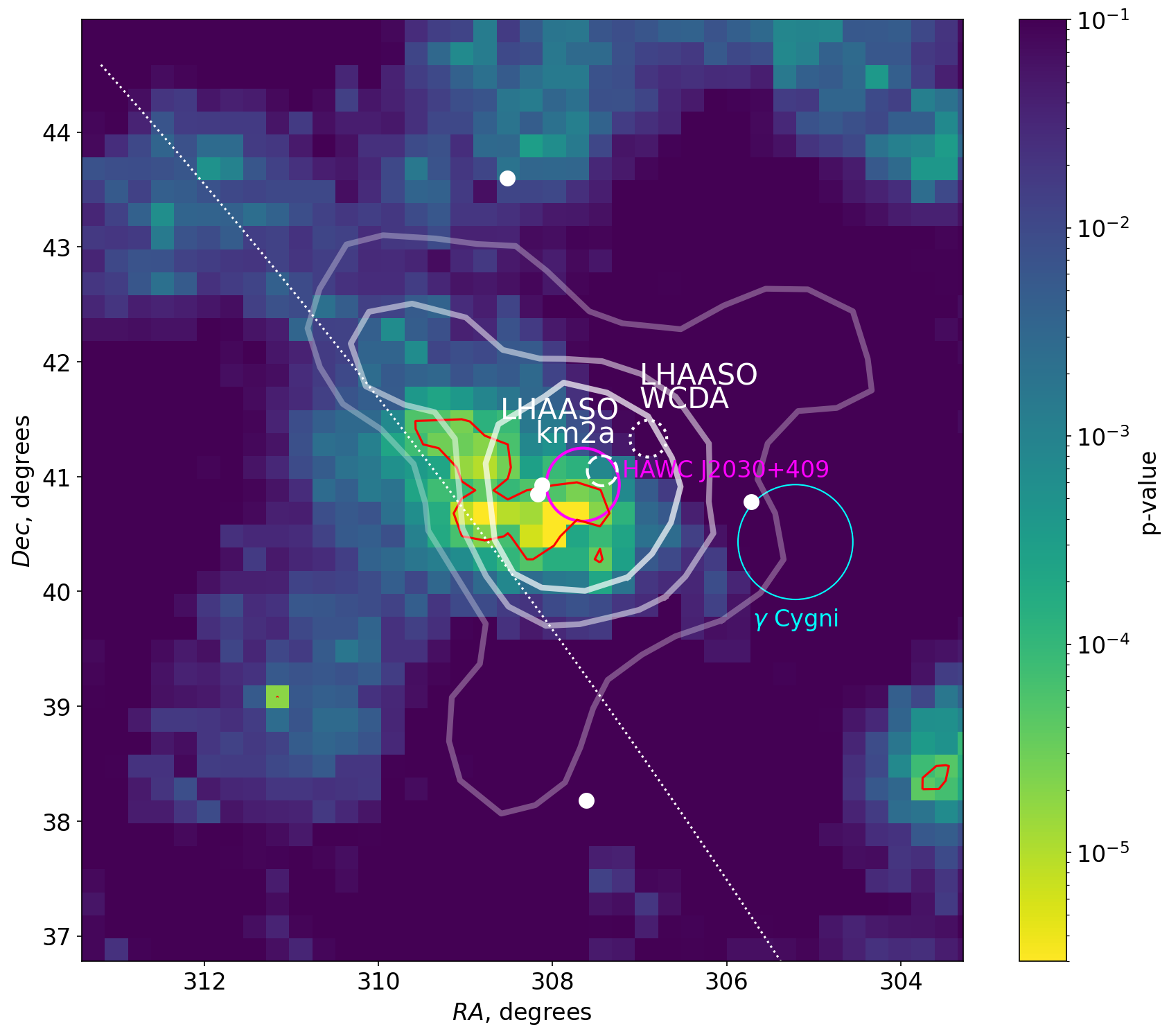}
    \includegraphics[width=0.9\linewidth]{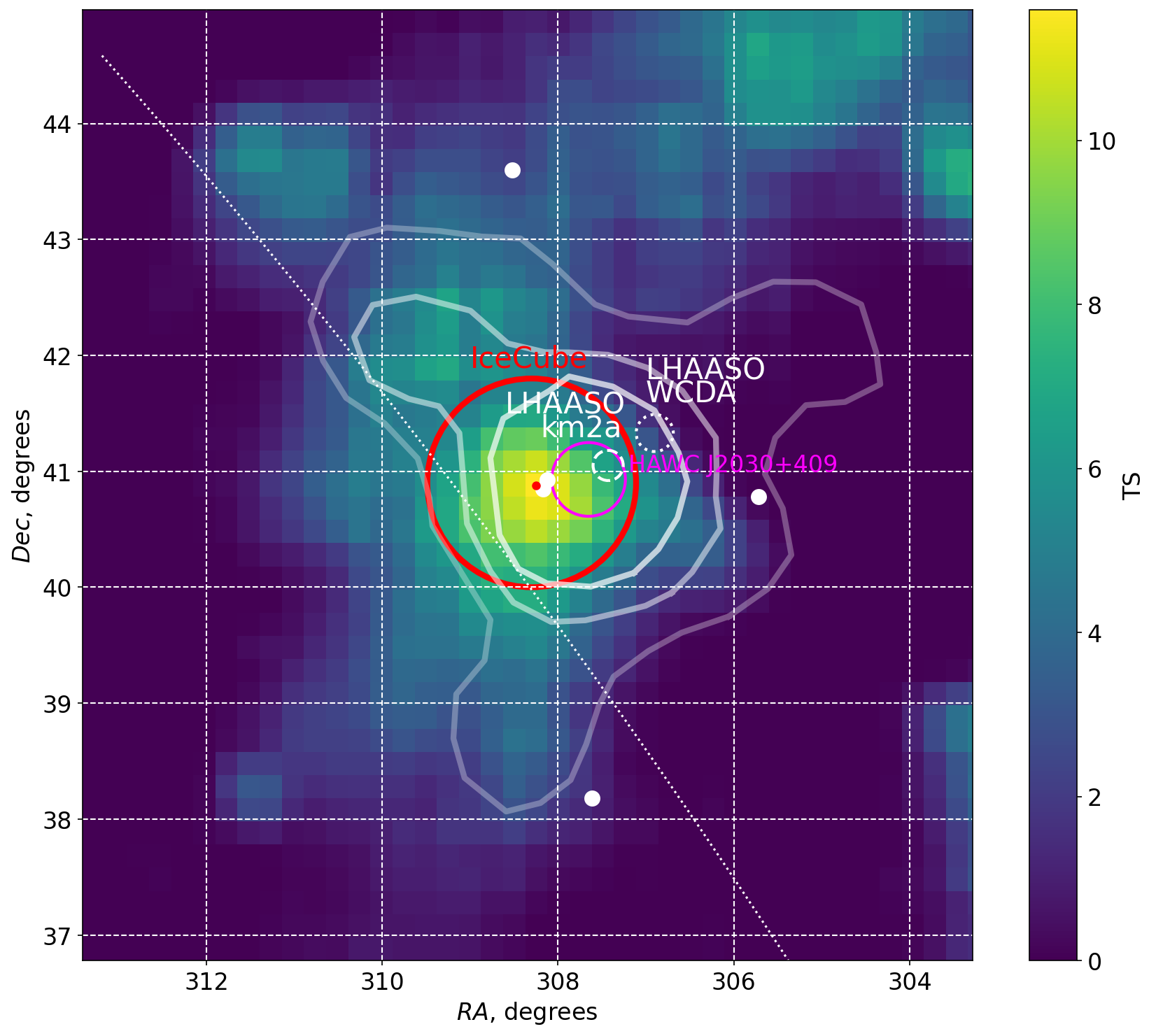}
    \caption{Top: map of  the $p$-values of (in)consistency of the count statistics with background-only hypothesis in the event counting analysis.  Cyan circle is the position of $\gamma$ Cygni supernova remnant. Magenta circle shows the uncertainty of localisation of the HAWC Cocoon source.  White dashed circle shows the uncertainty of position of LHAASO {KM2A} source, dotted circle is the position of the \gr\ source centroid measured by LHAASO WCDA. White points show the PeV energy \gr s detected by LHAASO {KM2A}. Contours show LHAASO significance contours of $12\sigma$, $15\sigma$, $18\sigma$ \cite{LHAASO:2023pum}. Inclined line marks the Galactic Plane. Red solid contour shows the $p_0=6\times 10^{-4}$ p-value level.  Bottom: 
    map of the TS values of the likelihood analysis. Red dot and circle show the bestfit and uncertainty of the position of the neutrino source. Other notations are the same as in the top panel.}
    \label{fig:image}
\end{figure}

The \gr\ data \cite{cocoon_fermi,hawc,argo,LHAASO:2023pum} indicate that the high-energy source in the Cygnus region has complex morphology. Fermi/LAT \cite{cocoon_fermi}, ARGO \cite{argo}, HAWC \cite{hawc} and LHAASO \cite{LHAASO:2023pum} telescopes all find that the signal is not from a point source. Modelling of the \gr\ signal morphology reveals a more compact source  (TeV J2032+4130, first detected by HEGRA telescope \cite{hegra}) close to the location of the pulsar PSR J2032+4127, and a more extended source associated with diffuse emission from a ``Cocoon" with radius about 2 degrees (first discovered by Fermi/LAT \cite{cocoon_fermi}). The overall \gr\ flux from the region is dominated by the Cocoon {\cite{hawc}}. 
This source, detected up to PeV energy range by LHAASO \cite{LHAASO:2023pum}, seems to have energy-dependent morphology: the position of the higher energy source seen by the KM2A detector of LHAASO is shifted with respect to that of the lower-energy WCDA detector source while a doublet  highest energy PeV photons is further displaced with respect to the position of the KM2A source centroid \cite{LHAASO:2023pum}, see Fig. \ref{fig:image}. The spectrum of the \gr\ source is softening with energy. LHAASO WCDA data are consistent with a powerlaw spectral model $dN/dE\propto E^{\Gamma}$ with the slope $\Gamma=-2.63\pm 0.08$, while the KM2A source is described by a powerlaw with the slope $\Gamma=-2.99\pm 0.07$ \cite{LHAASO:2023pum}. {Even though it is not possible to unambiguously determine the origin of the \gr\ emission,} HAWC \cite{hawc} and LHAASO \cite{LHAASO:2023pum} data interpretation suggest that the bulk of the \gr\ flux is of hadronic origin and thus has to have the neutrino counterpart.

\begin{figure}
   \includegraphics[width=0.95\linewidth]{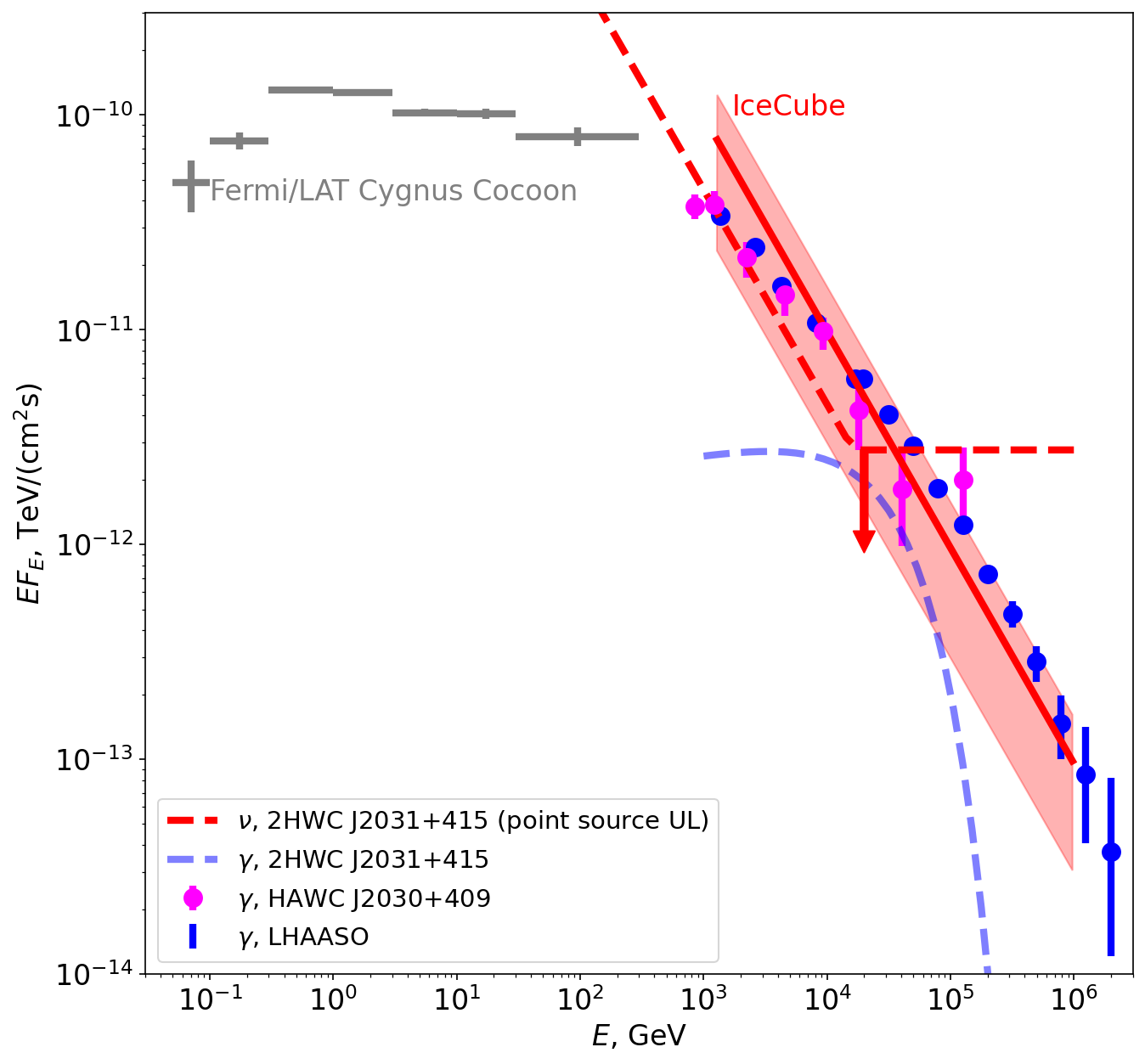}
    \caption{Spectrum of extended multi-messenger signal from the Cygnus region. Red solid line shows the best-fit all-flavour neutrino spectrum, red shaded region shows the  $68\%$ confidence interval of the neutrino flux level. The flux upper limits on the point source 2HWC J2031+415 from IceCube point source analysis {for the powerlaw spectra with slopes $\Gamma=2$ and $\Gamma=3$ are shown by the {red} dashed lines with an arrow \cite{icecube_10yr_paper}}. Bright blue data points show the measurements of the $\gamma$-ray flux from the extended Cocoon source  by LHAASO \cite{LHAASO:2023pum}. Magenta data points show HAWC spectrum of the Cocoon \cite{hawc}.  Grey data points show the Cygnus Cocoon \gr\ spectrum from the Fermi/LAT catalog \cite{fermi-cat}. {Pale blue dashed line shows a fit to \gr\ spectrum of the compact source in 2HWC J2031+415 from {Ref.} \cite{hawc}.} }
    \label{fig:sed}
\end{figure}

Search for the muon neutrino signal from the Cygnus region based on 10~year IceCube exposure \cite{icecube_10yr_paper} has only resulted in an upper limit {on a point-like source 2HWC J2031+415}. {The limit was derived for a powerlaw spectrum $dN_\nu/dE_{\nu_\mu}=F_0\left(E_{\nu_\mu}/1\mbox{ TeV}\right)^{\Gamma}$ with normalisation $F_0$ in $E_{\nu_\mu}\gtrsim 1$~TeV muon neutrino energy range. This limit, reported for $\Gamma=-2$ and $\Gamma=-3$ in Ref. \cite{icecube_10yr_paper}\footnote{{The analysis has identified a mild excess in the direction of the source, with the best-fit powerlaw spectrum $\Gamma=3.8$, but the flux estimate for such slope has not been reported.}}, is shown in Fig. \ref{fig:sed} as a limit on total neutrino flux\footnote{We assume total mixing of neutrino flavours, so that the total neutrino flux is 3 times higher than the muon neutrino flux.}. The neutrino flux limit is higher than the flux of the  compact \gr\ source at the position 2HWC J2031+415. However, it is comparable to the \gr\ flux measured by HAWC and LHAASO from the Cygnus Cocoon region, also shown in Fig. \ref{fig:sed}}. Within the hadronic model of the Cocoon activity, the neutrino flux is expected to be approximately equal to the \gr\ flux. {Closeness of the IceCube limit (on the point source flux) to the \gr\ source flux measurement suggests that the neutrino flux from an extended hadronic source may actually be detectable}. 

\textbf{Neutrino source in the direction of the Cygnus region.}
We use the publicly available dataset of IceCube  \cite{icecube_10yr_paper,icecube_10yr_data} to show that the neutrino signal from the Cygnus region is consistent with the expectations from the hadronic model. The main difference of our analysis from the approach of Refs. \cite{icecube_10yr_paper,icecube_10yr_data} {is that we look for the flux of extended, rather than point, source in the Cygnus region}. Similarly to Ref. \cite{IceCube_1068,Neronov:2023aks}, we consider the data of the full IceCube detector, because our methods rely on homogeneity of event selection.  {We use two complementary approaches for the source search: the unbinned likelihood analysis and the aperture photometry. }

Within the aperture photometry approach, we perform a search for a localised excess of muon neutrino events within a circle of radius $R$. Ideally, this excess is best seen in events of the best quality of angular reconstruction, $\sigma$, but the number of best quality events with {$\sigma\ll R$} is small and hence a trade-off between better localisation and larger signal statistics has to be made: {events with $\sigma\gtrsim R$ can be considered, but events with $\sigma\gg R$ would rather add background noise without increasing signal statistics. There exists an optimal choice of event sample with $\sigma<\sigma_{cut}$ for which the signal-to-noise ratio is maximal. }

We find an excess of events with the $p_0$  values between $6\times 10^{-6}$ and $8\times 10^{-5}$ for any possible source center location within the position uncertainty of HAWC or LHAASO source centroid. The number of the excess events in the signal region ranges between $99$ and $131$, depending on the source positioning, while the expected background ranges between $66$ and $91$ events. Top panel of Fig. \ref{fig:image} shows the map of the lowest  $p_{0}$ values for  event selections with different $\sigma_{cut}$, $0.5^\circ<\sigma_{cut}<2^\circ$ and  $0.5^\circ<R<2^\circ$ centered at the position of each pixel. The red solid contour shows $p_0=6\times 10^{-5}$ level, corresponding to the $4\sigma$ local significance of inconsistency of the count statistics  with the expectation from the background.  

The probability $p_{0}$ does not take into account the trial factor related to our choice of $R, \sigma_{cut}$. To calculate this factor, we perform Monte-Carlo simulations, assigning random Right Ascensions to the IceCube events while leaving their declinations unchanged.  We repeat the event counting analysis for each simulated data set, finding the best $R, \sigma_{cut}$ minimising the $p$-value. Counting the number of occurrences of the simulated data sets with fluctuations of event statistics resulting in $p$-values at least as as low as $p_{0}$, we estimate the post-trial probability for the observed excess to be a background fluctuation better than $p=2.7\times 10^{-3}$, i.e. source detection significance at $\gtrsim 3\sigma$ level for any source center position. 

For the likelihood analysis, we calculate the unbinned likelihood \cite{mattox96}
\begin{equation}
\label{eq:likelihood}
\log L(N_s)=\sum_i\log\left(\frac{N_s}{N_t}S_i+\left(1-\frac{N_s}{N_t}\right)B_i\right)
\end{equation}
where the sum runs over the neutrino event sample,  $N_s,N_t$ are the source and total event counts, \linebreak $S_i$ and $B_i$ are the  probability density functions  (PDF) of the signal and background for $i$-th event. 

The background on top of which the source signal is detected depends on the declination, because of the geographical location of IceCube at the South Pole. Its spectral and spatial PDF is determined by the atmospheric neutrinos \cite{2011PhRvD..83a2001A}. Assuming that the source signal provides a minor contribution to the overall count statistics, we calculate $B$ directly from the data, by computing the distribution of detected events in declination and energy.  

{The source PDF depends on the assumed shape of the source spectrum and on the parameter(s) of the spatial model. Similarly to previous IceCube source searches \cite{icecube_10yr_paper,2019ApJ...886...12A}, we consider the powerlaw spectral models with fixed slope $\Gamma=-3$  consistent with the LHAASO KM2A  \gr\ data. The spatial model is characterised by one parameter, $R$, the source extension. The point source hypothesis considered in Ref. \cite{icecube_10yr_paper} corresponds to $R=0$.   We consider two types of spatial models: a flat disk 
\begin{equation}
{\cal M}(\vec r\, ',R)=\left\{\begin{array}{ll}
1/(\pi R^2), &|\vec r\, '|<R\\
0, &|\vec r\, '|>R
\end{array}
\right.
\end{equation}
where $\vec r$ is the angular displacement of event  from the source center position  and $R$ is the disk radius, or a two-dimensional Gaussian of the width $R$:
\begin{equation}
{\cal M}(\vec r\, ',R)=\frac{1}{2\pi R^2}\exp\left(-\frac{\vec r\, '^2}{2R^2}\right)
\end{equation}
Some of the lower quality IceCube events can have insufficient precision of angular reconstruction $\sigma$, that is comparable or larger than the source size we are interested in ($\sim 2^\circ$). We exclude such events from out analysis by imposing a quality cut $\sigma<1^\circ$. We have verified that such a cut does not degrade the outcomes of the analysis, checking that exclusion of such low-quality events does not degrade the TS value for the  sources for which an evidence of the signal has been previously reported by IceCube collaboration \cite{IceCube_1068,txs,txs1}. We assume that the point spread function for the selected events with good quality of reconstruction of the angular direction is well described by a two-dimensional circularly symmetric  Gaussian. To calculate $S$ we convolve the spatial models of the source with  two-dimensional Gaussian  kernel $G_\sigma(\vec r,\vec r\, ')$ of the width equal to the angular error $\sigma$:
\begin{equation}
S(\vec r)=\int {\cal M}(\vec r\, ')G(\vec r,\vec r\, ')d^2\vec r\, '
\end{equation}
The functions $S$ are tabulated over a  $10^\circ\times 10^\circ$ ROI. 

{To test the presence of the signal, we calculate the Test Statistic $TS(N_s)=2(\log L(N_s)-\log L(0))$
that compares the likelihood of the presence of non-zero signal with any number of counts $N_s$ to the likelihood of the null hypothesis of zero signal counts. Bottom panel of Fig. \ref{fig:image} shows the TS value map of the Cocoon region.   
We find a disk (Gaussian) source with $TS=11.6$ ($10.8$) for the extension $R=1.2^\circ$ ($0.7^\circ$),  at the position of $RA=208.3^\circ\pm 0.9^\circ$,\ Dec=$40.9^\circ\pm 0.9^\circ$ provides the best fit to the data. Red dot and circle in the bottom panel of Fig. \ref{fig:image} show the bestfit source position and its uncertainty. Fig. \ref{fig:TS_chart} shows the dependence of the TS value on $R$.  One can see that only a mild excess of $N_s<20$ events is seen for $R=0$. This is consistent with the analysis of the point source at the nearby position of 2HWC J2031+415 in which an excess of 13 events is found \cite{icecube_10yr_paper}. 

Given the limited statistics of the neutrino data we perform the spectral analysis fixing the spectral slope to $\Gamma=-3$ consistent with LHAASO KM2A measurement. To convert the physical source flux in the neutrino count number, we use the instrument response functions of IceCube \cite{icecube_10yr_data}. We calculate the expected energy distribution of detectable neutrinos in narrow  energy intervals $dE_{\nu_\mu}$ by multiplying the neutrino spectrum $dN_{\nu_\mu}/dE_{\nu_\mu}$ by the neutrino effective area $A_{\nu_\mu}(E_{\nu_\mu})$ and exposure time  $T_{exp}$ that we find summing all time intervals in the ``uptime" tables provided with the data release \cite{icecube_10yr_data}.
We calculate the expected muon energy distribution by convolving the neutrino energy distribution with the ``smearing matrices"  that provide the probability density functions  to find muon events with given reconstructed muon energy $E_\mu$, angular reconstruction precision $\sigma$ and misalignment with respect to the reference neutrino arrival direction, $\theta$, for given neutrino energy $E_{\nu_\mu}$ and declination $Dec$. To find the flux normalisation we fit the muon counts in each energy bin using 
likelihood defined through the W-statistic (we use implementation from the \texttt{gammapy} package \cite{gammapy:2017,gammapy:2019}). 
The flux normalisation found in the aperture photometry analysis is 
$F_0=3.3_{-1.0}^{+2.0}\times 10^{-11}/\mbox{(TeV cm}^2\mbox{s)}$.
In the likelihood analysis we estimate the flux uncertainty as an interval of TS values larger than $TS_d-4.5$, given that we adjust  disk size $R$ along with  $N_s$ \cite{wilks38}. This gives $
    F_0=3.3_{-2.3}^{+2.1}\times 10^{-11}/\mbox{(TeV cm}^2\mbox{s)}$
consistent with the result of the aperture photometry analysis. Red solid line and red error band in the Fig. \ref{fig:sed} show the estimate of the disk source flux from the likelihood analysis. We show the all-flavour neutrino flux, that is directly comparable to the \gr\ flux within the hadronic models. We assume the full mixing of the neutrino flavours.

\begin{figure}
    \centering
    \includegraphics[width=0.9\linewidth]{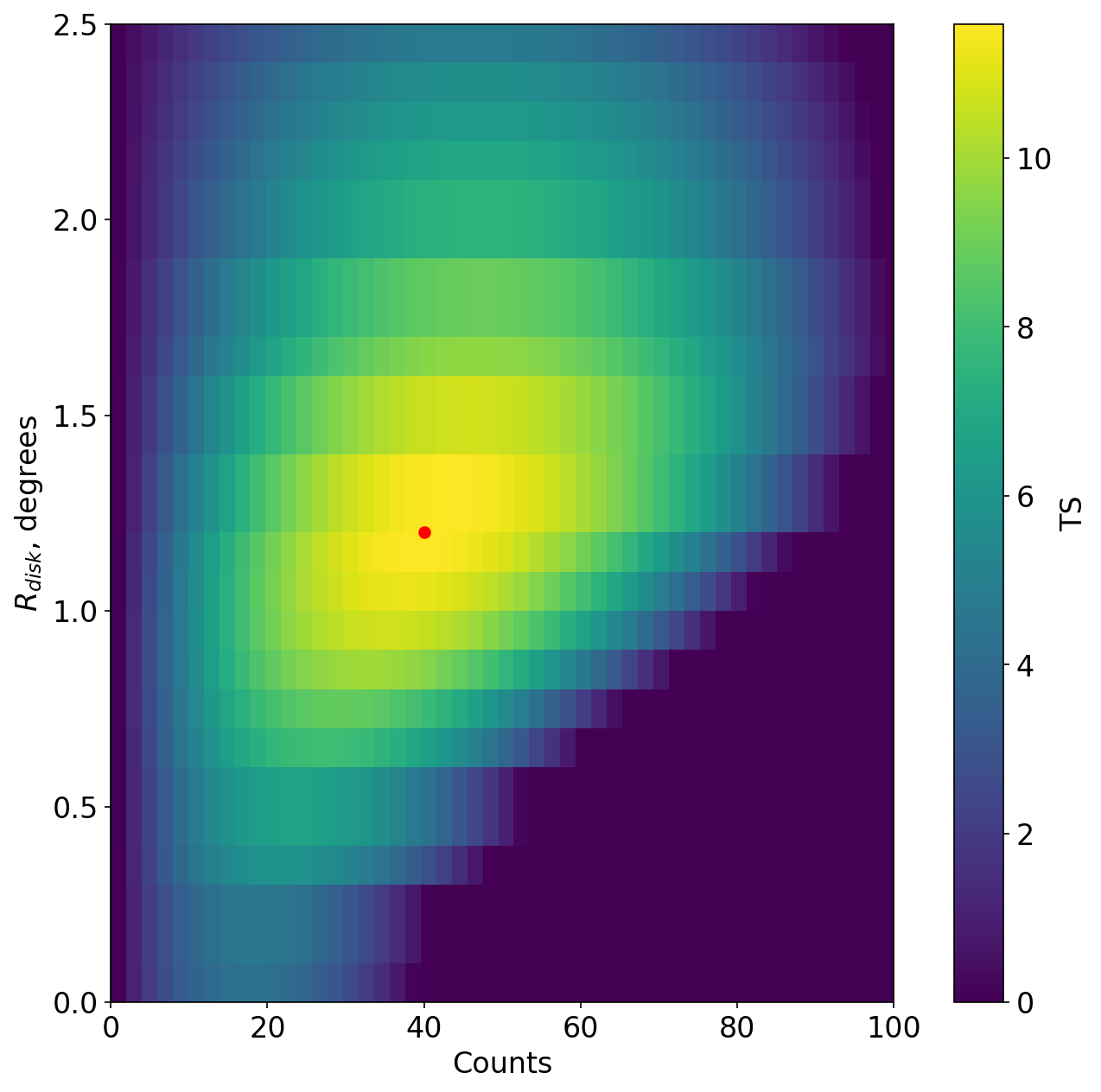}
    \caption{{Dependence of TS value on $R$ for the disk source. Red dot shows the best fit count number and source extension. }}
    \label{fig:TS_chart}
\end{figure}

\textbf{Discussion and conclusions.}
We searched for the extended neutrino excess in the direction of Cygnus region in the publicly available ten-year IceCube neutrino telescope dataset. We have found an excess of events from the direction of Cygnus Cocoon inconsistent with a backgorund fluctuation at $\simeq 3\sigma$ confidence level.  The neutrino flux from the source  is found to be comparable to the flux of \gr s in the multi-TeV energy range seen by HAWC and LHAASO.

 {Our analysis confirms the hypothesis that the emission from the Cygnus Cocoon region is powered by interactions of cosmic ray protons and atomic nuclei. It reveals an evidence for a} Galactic high-energy neutrino  source on the sky. Low statistics of the  neutrino data does not allow us to see if there is a compact (or even point-like) neutrino source at the location of either PSR J2032+4127, detected by HAWC and LHAASO (HAWC J2031+415 or LHAASO J2032+4130) or at Cyg X-3 that has been considered as candidate neutrino source based on \gr\ data \cite{1986ApJ...301..235B,2014ApJ...780...29S}.   The complex morphology of the extended multi-messenger source may be influenced not only by the location of possible point(s) of injection of  cosmic rays, but also by the effect of anisotropic diffusion of cosmic rays preferentially spreading along the direction of the ordered Galactic magnetic field, which is almost aligned along the line-of-sight in the Cygnus direction \cite{jansson12,giacinti17}. Better constraints on the properties of the neutrino source  will be possible with the increase of the neutrino signal statistics provided by the new km$^3$ class neutrino telescopes KM3NET \cite{km3net} {and} IceCube-Gen2 \cite{2021JPhG...48f0501A}.

\bibliographystyle{Science}
\bibliography{refs}

\end{document}